\documentclass[secnumarabic,amssymb,nobibnotes, aps, prd,12pt]{revtex4-2}

\usepackage{multirow}
\usepackage{amsmath,amsfonts}
\usepackage{amssymb,latexsym}
\usepackage{color}
\usepackage{graphicx}
\usepackage{mathrsfs}
\usepackage{physics}

\usepackage{mathtools}

\usepackage{graphicx}% Include figure files
\usepackage{dcolumn}% Align table columns on decimal point
\usepackage{bm}

\usepackage{amsmath,amsthm,amsfonts}
\usepackage{latexsym}
\usepackage{stmaryrd}
\usepackage{color}
\usepackage[applemac]{inputenc} 
\usepackage{fancyhdr}
\usepackage[T1]{fontenc}
\usepackage{dcolumn}% Align table columns on decimal point
\usepackage{soul}

\usepackage{epsfig}
\usepackage{epstopdf}
\usepackage[all]{xy}
\usepackage[unicode=false]{hyperref}
\usepackage{url}
\usepackage{dsfont}
\usepackage{slashed}
\usepackage{mathrsfs}

\usepackage{bbm}
\usepackage{tikz}
\usepackage{physics}
\usepackage{enumitem}
\usepackage{float}
\usepackage{silence}
\usepackage{float}

\newtheorem{prop}{Proposition}
\newtheorem{assum}{Assumption}

\newcommand{\beprop}{\begin{prop}}
\newcommand{\enprop}{\end{prop}}
\newcommand{\bprf}{\begin{proof}} 
\newcommand{\eprf}{\end{proof}\qed}

\definecolor{hervecolor}{rgb}{0.8,0,0.7}

\newcommand{\scalar}[2]{\langle\kern.3ex #1 \kern.3ex|\kern.3ex#2\kern.3ex\rangle}

\newcommand{\ii}{\mathsf{i}}

\newcommand{\by}{\mathbf{y}}
\newcommand{\bx}{\mathbf{x}}
\newcommand{\br}{\mathbf{r}}
\newcommand{\bq}{\mathbf{q}}
\newcommand{\bp}{\mathbf{p}}
\newcommand{\bb}{\mathbf{b}}
\newcommand{\bQ}{\mathbf{Q}}
\newcommand{\bA}{\mathbf{A}}
\newcommand{\bP}{\mathbf{P}}
\newcommand{\bu}{\mathbf{u}}

\newcommand{\bv}{\mathbf{v}}
\newcommand{\calH}{{\mathcal H}}

\newcommand{\R}{\mathbb{R}}

\newcommand{\C}{\mathbb{C}}
\newcommand{\Z}{\mathbb{Z}}
\def\lg{\langle }
\newcommand{\rg}{\rangle }

\newcommand{\vap}{\varpi}

\newcommand{\ud}{\mathrm{d}}

\newcommand{\sfM}{\mathsf{M}}

\newcommand{\sfMv}{\mathsf{M}^{\vap}}

\newcommand{\Om}{\Omega}

\newcommand{\cdm}{\mathsf{C}_{\mathrm{DM}}}

\newcommand{\bun}{\mathbbm{1}}

\begin{document}
\numberwithin{equation}{section}
\title{A purely geometrical Aharonov-Bohm effect}

\author{Jean-Pierre Gazeau}
\email{gazeau@apc.in2p3.fr,j.gazeau@uwb.edu.pl}
\affiliation{Universit\'e Paris Cit\'{e}, CNRS, Astroparticule et Cosmologie, F-75013 Paris, France}
\affiliation{Faculty of Mathematics, University  of Bia\l ystok, 15-245 Bia\l ystok, Poland}
\author{Tomoi Koide}
\email{tomoikoide@gmail.com, koide@if.ufrj.br}
\affiliation{Instituto de F\'{i}sica, Universidade Federal do Rio de Janeiro, Rio de Janeiro, RJ, Brazil}
\author{Romain Murenzi}
\email{rmurenzi@wpi.edu}
\affiliation{Dep. Physics, Worcester Polytechnic Institute, Worcester, MA 01609, USA }
\author{Aidan Zlotak}
\email{ahzlotak@wpi.edu}
\affiliation{Dep. Physics, Worcester Polytechnic Institute, Worcester, MA 01609, USA }

\date{\today}

\leavevmode
\begin{abstract}
 We present an application of the affine covariant integral quantization (ACIQ) (Adv. Oper. Theory, 5, 2020; Adv. Oper. Theory, 7, 2022) to quantum mechanics on the punctured  plane. The associated four-dimensional phase space is identified with the similitude group SIM(2), which comprises translations, rotations, and dilations of the plane. Due to the topology of the punctured  plane, our quantization procedure gives rise to an affine vector potential. This potential can be interpreted as the Aharonov-Bohm (AB) gauge field produced by an infinite solenoid. This observation supports a reinterpretation of the AB effect: it emerges from the topological constraint imposed by the impenetrable coil rather than from an externally applied classical gauge field. In addition to this gauge structure, ACIQ also generates a repulsive, centrifugal-like scalar potential, a feature already encountered when applying ACIQ to motion on the half-line, whose phase space is the open half-plane. These results provide a new perspective on the AB effect, highlighting the central roles of topology and symmetry in quantum mechanics.
\end{abstract}

\maketitle

\tableofcontents
\section{Introduction }
\label{intro}

The Aharonov-Bohm (AB) effect is a cornerstone of quantum mechanics, illustrating how topology influences physical observables \cite{ahabohm59}.
Despite experimental verification, its deeper implications, especially concerning quantization and gauge potentials, remain under debate. Its background is summarized, for example, 
in the introductions of recent papers by Aharonov and other research groups \cite{aharo16,choen23}. Our work focuses specifically on the issues of the AB effect from the perspective of quantization.

One criticism of the existence of the AB effect is that the gauge potential in the Schr\"{o}dinger equation is merely apparent and that the equation can be reformulated entirely in terms of the electric and magnetic field's observables. Based on this view, Strocchi and Wightman used the so-called Madelung transformation to rewrite the Schr\"{o}dinger equation as a hydrodynamic equation \cite{wightman74}. In this hydrodynamic form, all gauge potentials are replaced by electromagnetic fields, leading them to argue that the AB effect is impossible. 
A similar criticism has been raised by DeWitt \cite{dewitt}.

However, as is well known, the hydrodynamic representation of the Schr\"{o}dinger equation is not always equivalent to the Schr\"{o}dinger equation itself. For example, in cases involving stationary states under central potentials where the wave function has nodes, the quantum potential term in the hydrodynamic equation becomes singular. In such situations, as pointed out by Bohm \cite{bohm79}, additional geometric conditions are necessary to ensure the single-valuedness of the wave function, and it is through these conditions that the gauge potential appears.

A similar phenomenon occurs in operator formalism. As discussed in Appendix \ref{app:canonical}, introducing a velocity operator instead of the momentum operator allows the elimination of gauge potentials from both the Schr\"{o}dinger equation and the canonical commutation relations. However, constructing operators that satisfy these commutation relations inevitably requires introducing degrees of freedom corresponding to gauge potentials. Thus, it is impossible to ignore the role of gauge potentials within the framework of quantum mechanics.

Since Tonomura's experiment \cite{tonomura83}, few have doubted the existence of the AB effect. However, its physical implications remain a subject of ongoing debate. For instance, recent efforts have been made to reinterpret the AB effect from the perspective of information theory \cite{choen23}.

In practical treatments of the AB effect, it is common to consider a situation where charged particles, say  electrons, are prevented from entering the interior of the coil, thereby avoiding direct interaction with the magnetic field. This restriction limits the allowed electron motion and consequently alters the definition of self-adjointness. Standard canonical quantization is based on the translational symmetry of the phase space, and the canonical commutation relation (CCR) ( $[\hat{q},\hat{p}]=i \hbar \bun_{\mathrm{d}}$ ) holds for a pair of self-adjoint operators, is generally applicable only when the spectra of these operators are the whole real line, with certain notable exceptions \cite{schmudgen83,schmudgen12}. In other words, a new quantization scheme incorporating topological effects is needed to construct quantum mechanics for systems without translational symmetry. Thus, the interference effects in the wave function's phase observed in the AB effect may be attributed to topological effects rather than to the gauge potential generated by the coil.

Among the various explanations of the AB effect, some studies have attributed its origin to the influence of topology on the quantization of canonical variables. For example, Ohnuki and collaborators extended canonical quantization to systems on a spherical surface, showing that topological effects can naturally induce gauge potentials as a consequence of quantization \cite{ohnuki93,landsman91,ogawa97,ohnukisis95,ohnuki97,tsutsui94,tsutsui95}. However, an established scheme for topological quantization is still lacking, Ohnuki et al., for instance, consider particle motion on hypersurfaces. To more clearly elucidate the relationship between topology and the AB effect, it is necessary to consider quantization in the case of a flat plane that includes a singular region which the wave function cannot penetrate. The behavior of electrons around an infinitely long coil has been studied as an idealized model of the AB effect. If electrons cannot penetrate the interior of the coil, that interior becomes a spatial singularity. For further simplicity, one may consider a plane perpendicular to the coil and take the limit of an infinitesimal coil radius, effectively yielding a plane with a singularity. If topological quantization explains the AB effect, it should naturally produce a gauge potential around the singularity consistent with classical electromagnetism.

To study this system, we employ the affine covariant integral quantization (ACIQ) method \cite{gakomu20,gakomu22} on a plane with singularities. ACIQ serves as a natural alternative to canonical quantization when the configuration space lacks translational invariance but instead exhibits dilation symmetry. This approach has also demonstrated its effectiveness in other areas, notably in quantum cosmology (see \cite{bergeron24} and references therein; see also the pedagogical review in \cite{almeida18}). By applying ACIQ to the Aharonov-Bohm (AB) effect, we demonstrate for the first time that the quantization procedure naturally generates a gauge vector potential analogous to that produced by an infinite solenoid. This potential arises from topological effects associated with the singularity, revealing the AB effect as an emergent feature of the underlying quantum symmetry. In addition to this gauge structure, ACIQ also induces a repulsive, centrifugal-like, scalar potential - a characteristic already observed when ACIQ is applied to motion on the half-line, where the corresponding phase space is the open half-plane \cite{bergeron24,almeida18}.

Let us clarify the term \textit{affine covariant integral quantization}. The adjective ``affine'' pertains to the group of symmetries of the phase space, encompassing both translations and dilations (and rotations in the present context),  a symmetry distinct from the translational symmetry underlying canonical or Weyl-Heisenberg quantization. ``Covariant'' indicates that the quantization map intertwines classical (geometric operations) and quantum (unitary transformations) symmetries. ``Integral'' signifies the use of integral calculus as a fundamental tool in implementing the method, particularly when applied to singular functions or distributions, where integral techniques are indispensable. Coherent state quantizations, also referred to as Berezin-Klauder-Toeplitz quantizations, can be viewed as particular cases within the broader class of integral quantizations.

This paper is structured as follows:
In Sec.\ \ref{solmagrev}, we review textbook material on the magnetic and vector potentials generated by an infinite solenoid, first considering a finite radius and then the limiting case of a zero-radius solenoid. The resulting Hamiltonian for a charged point particle moving on the punctured plane is also derived.
In Sec.\ \ref{2daffG}, we outline the group structure of the relevant phase space $\R^2_\ast\times \R^2$, where  $\R^2_\ast\equiv \R^2\setminus\{\mathbf{0}\}$ denotes the plane with the origin removed.
In Sec.\ \ref{2dcovaffq}, the affine covariant integral quantization (ACIQ) method is introduced, with specific examples relevant to the present study.
In Sec.\ \ref{geompot}, we present the explicit forms of the geometric vector and scalar potentials derived from affine quantization and discuss their semi-classical portraits. 
This material is then applied to the Aharonov-Bohm (AB) model, highlighting several key findings.
In Sec.\ \ref{disc}, the concluding remarks are provided, summarizing the main results and potential implications.
The technical aspects of our approach are detailed in the Appendices.
In Appendix \ref{app:canonical}, basic considerations regarding canonical quantization.
In Appendix \ref{app:examples=quantization}, practical examples of ACIQ, emphasizing the most relevant cases for the present work.
In Appendix \ref{rankone}, the specific case of ACIQ with density rank-one operators, i.e. affine coherent states, is examined.
In Appendix \ref{app:formulae}, the mathematical formulae necessary for calculating the gauge potential  induced by topology are summarized.

To conclude this introduction and before delving into the core content of this work, it is essential to clarify the rationale and structure of some of the forthcoming sections. Sections \ref{2daffG} and \ref{2dcovaffq}, along with Appendices \ref{app:examples=quantization} and \ref{rankone}, provide a summary of the mathematical framework underlying the  affine  covariant integral quantization (ACIQ) method. This material is drawn and adapted from our prior works \cite{gakomu20, gakomu22}, originally presented in a mathematically oriented journal. We have restructured these foundational elements with a focus on the needs of a physics-oriented readership, emphasizing the physical implications and technical nuances critical to our study.

\section{Magnetic field and potential due to infinite solenoid}
\label{solmagrev}

Let us revisit the basic material dealing with an infinite coil of radius $a$ aligned along a unit vector ${\bf e}_3$. The motion of a particle of mass $m$ and charge $\mathfrak q$ is confined to the ${\bf e}_3$-plane, perpendicular to the coil axis. A magnetic field $\mathbf{B}( \br )$ is formed by the coil which has the current density  
\begin{eqnarray}
\label{curj}
{\bf j} = n I \delta(r-a) {\bf e}_\theta \, ,
\end{eqnarray}
where $I$ is the current, $\delta(\cdot)$ is the Dirac function, and $n$ is the number of turns.
Let $({\bf e}_1,{\bf e}_2,{\bf e}_3)$ form a direct orthonormal frame. The corresponding  orthonormal frame in cylindrical coordinates is denoted by $({\bf e}_r,{\bf e}_\theta,{\bf e}_3)$. 

It is well-known that the current density \eqref{curj} induces the static magnetic field which gives a finite contribution only inside the coil, forming a vector potential parallel to the normal vector of the plane.
For the sake of simplicity, we consider the infinitesimal limit of the radius of the coil $a \rightarrow 0+$.
Assuming that the magnetic field ${\bf B}$ is finite at the origin, $(x_1,x_2) = (0,0)$ in  Cartesian coordinates, the static Maxwell equations give the following gauge potentials:  
\begin{equation}
\label{vecfield}
{\bf A}^{sol} = A^{\mathrm{sol}}_{x_1} {\bf e}_1 + A^{\mathrm{sol}}_{x_2} {\bf e}_2 \, ,
\end{equation}
where
\begin{equation}
A^{\mathrm{sol}}_{x_1}= -\frac{\Phi_0}{2\pi} \, \frac{x_2}{{r}^2}\, , \quad      A^{\mathrm{sol}}_{x_2}= \frac{\Phi_0}{2\pi} \, \frac{x_1}{{r}^2} \, .
\end{equation}
with $r$ being $\sqrt{x^2_1 + x^2_2}$. 
We consider a finite magnetic flux $\Phi_0$, which remains dependent on $n$, $J$ and $a$, even  as $a$ approaches zero.

The corresponding Hamiltonian is
\begin{equation}
\label{exclham}
H_0( \bx  ,  \bp  ) =  \frac{1}{2m}\left( \bp  - \mathfrak{q}\,\mathbf{A}^{\mathrm{sol}}\right)^2=  \frac{p^2}{2m}- \frac{\mathfrak{q}}{m}\,\frac{\Phi_0}{2\pi x^2} \bx\wedge\bp + \frac{1}{2m}
\left(\frac{\mathfrak{q}\Phi_0}{2\pi }\right)^2 \, \frac{1}{x^2}\,, \quad x=\Vert\bx\Vert\,. 
\end{equation}
Here $({\bx}, {\bp} )$ denotes a pair of canonical variables.
Due to the presence of the infinitely thin solenoid, the origin of the plane has to be considered as a singularity, and the classical phase space is defined by $\Gamma\equiv\mathbb{R}_{\ast}^2\times \mathbb{R}^2=\{( \bx , \bp )\, | \, \bx \in \mathbb{R}_{*}^2\,, \, \bp \in\mathbb{R}^2\}$, \underline{not} just by $\mathbb{R}^4$. 
Consequently, the standard canonical quantization is not applicable in this context. The geometry of the phase space $\Gamma$ corresponds exactly to that of the similitude group acting on the plane, also referred to as the 2D affine group, SIM$(2)$. Thus, a new quantization scheme aligned with this geometric framework is necessary - a framework extensively developed in the previous work \cite{gakomu20} and its complementary study \cite{gakomu22}.

\section{2d-Affine group as a phase space} \label{2daffG}

\subsection{2D-Affine Group}

Henceforth, all quantities under consideration will be treated as dimensionless, and $\hbar = 1$.

Note that, in agreement with the notations used in the previous section, Cartesian coordinates of a vector $\bx$ in $\mathbb{R}_{\ast}^2$ or $\R^2$ will be denoted as $\bx= x_1 \mathbf{e}_1 + x_2 \mathbf{e}_2$.   
Then the group SIM$(2)$ is  formed of translations by $ \bb   \in \mathbb{R}^2$, dilations by $a \in \mathbb{R}^+_*:= \{x \in \mathbb{R}|\ x >0\}$, and rotations by $\theta \in [0,2\pi)$ mod $(2\pi)$. 
The action of an element $(a,\theta, \bb  )$
of SIM(2) on  $\mathbb{R}^2$ is given by:
\begin{equation}
\label{affact}
(a,\theta, \bb  )  \bx   = a\,\mathrm{R}(\theta) \bx  + \bb  \, ,
\end{equation}
where 
\begin{equation}
\mathrm{R}(\theta)= \begin{pmatrix}
 \cos\theta     & - \sin\theta   \\
  \sin\theta     &   \cos\theta
\end{pmatrix} \, ,
\end{equation}
is the rotation matrix in two dimensions acting on the column vector $\bx$.
The group composition law reads as,
\begin{equation}
(a,\theta, \bb)  (a^{\prime},\theta^{\prime}, \bb  ^{\prime})=(aa^{\prime},\theta+\theta^{\prime},a\,\mathrm{R}(\theta) \bb  ^{\prime}+ \bb  )\, , 
\end{equation}
and the inverse is given by
\begin{equation}
(a,\theta, \bb  )^{-1}=\left(\frac{1}{a}, -\theta, -\frac{\mathrm{R}(-\theta) \bb  }{a}\right)\, . 
\end{equation} 
 
Our phase space variables $( \bq , \bp )\in \Gamma$ for the motion of a particle in the punctured plane $\mathbb{R}_{*}^2=\{ \bx   \in \mathbb{R}^2\,  |\,   \bx    \neq \mathbf{0}\} $ are given in terms of $(a,\theta, \bb)$ as
\begin{equation}
\label{qpathtp}
q=\frac{1}{a}\, , \  \bq  = (q,\theta ) \,, \  \bb  \equiv \bp \, , 
\end{equation} 
where $(q,\theta)$ are polar coordinates for $\bq$. With this definition, it is important to note that $\bq$, as a scaling affine parameter.

\subsection{Complex algebraic structure of the phase space}

So far, we have considered scaling and momentum vectors $ \bq=(q_1,q_2) $ and $ \bp= (p_1,p_2) $ to denote an element  in the phase space $\Gamma \simeq$ SIM$(2)$.
To simplify the calculations, we introduce an equivalent complex notation, where $\bq=(q_1,q_2)\mapsto q_1+ \ii q_2 $ with its conjugate $\bq^* \mapsto q_1 - iq_2$.  This allows to define the multiplication rule 
\begin{equation}
\mathbf{q}\mathbf{ q}^{\prime}= (q_1q^{\prime}_1-q_2q^{\prime}_2,q_1q^{\prime}_2 +q_2q^{\prime}_1) \, .
\end{equation}
 It results that $\mathbf{e}_1 \mathbf{q} = \mathbf{q}$ for all $\mathbf{q}$, $\mathbf{e}_2 \mathbf{e}_2=-\mathbf{e}_1$, and so $\mathbf{e}_2\mathbf{q}=\mathbf{e}_2(q_1,q_2)= (-q_2,q_1)$.  One thus should avoid the confusion  between this ``imaginary'' $\mathbf{e}_2$ and  the ``$\ii$'' appearing in the canonical quantization $ \bp \mapsto \mathbf{ P}=-\ii \pmb{\nabla}$, where $\pmb{\nabla}=\left(\partial/\partial q_1, \partial/\partial q_2 \right)$ (reminder that  we set $\hbar=1$).
The Euclidean inner product is equally denoted by $\mathrm{Re}(\bq^\ast \bq')$ or  $\bq\cdot\bq^{\prime}$. 
Consistently, $\mathbf{e}_1\cdot \bq=q_1$ and $\mathbf{e}_2\cdot \bq=q_2$. 
Also note that we are  using the notation $\mathbf{e}_1=(1,0)$ or $1$ when there is no risk of confusion. For instance, we  occasionally write $\left.F(\bq)\right\vert_{\bq=\mathbf{e}_1}= F(1)$.

As already indicated in \eqref{qpathtp},  vectors $ \bq $ and $ \bp $ are biunivocally associated with the complex numbers $z( \bq )= qe^{\ii \theta}\equiv  \bq $ and $z( \bp )\equiv  \bp $ (same notation) respectively, and $\mathbf{1}\equiv 1$. Thus,  the multiplication of two vectors reads as
$
 \bq \, \bq^{\prime}=(q,\theta ) \,(q^\prime,\theta^\prime ) = (qq^{\prime}, \theta + \theta^{\prime}) 
$,
(in polar coordinates, with mod $2\pi$ for the angular coordinate), and for the inverse,
$
{ \bq }^{-1}=\left(1/q,-\theta\right )
$.
Multiplication and inverse in the SIM$(2)$ group read:
\begin{equation}
\label{afflaw}
( \bq , \bp )( \bq^{\prime}, \bp^{\prime})=\left( \bq  \bq^{\prime},\frac{ \bp^{\prime}}{ \bq^{\ast}}+ \bp \right)\, ,  \quad ( \bq , \bp )^{-1}=\left(\bq^{-1},- \bq  ^{\,\ast} \bp \right) \, , 
\end{equation}
where $ \bq  ^{\,\ast}=(q,-\theta ) \equiv \bar z( \bq )$. 
Hence, SIM$(2)$ is the semidirect product  of the two abelian groups $\R^+_\ast\times \mathrm{SO}(2)$ and $\R^2$:
\begin{equation}
\label{semidir}
\mathrm{SIM}(2)=  \left(\R^+_\ast\times \mathrm{SO}(2)\right)\ltimes\R^2 \sim \C_{\ast} \ltimes \C \,,
\end{equation}
where $\C_{\ast}=\C\setminus\{0\}$.
The left action of $\mathrm{SIM}(2)$ on itself is derived from \eqref{afflaw} as 
\begin{equation}
\label{leftaction}
\mathrm{SIM}(2) \ni ( \bq _0, \bp _0): \, ( \bq , \bp ) \mapsto ( \bq^{\prime}, \bp^{\prime})=( \bq _0, \bp _0)( \bq , \bp )=
\left( \bq _0 \bq ,\frac{\bp}{\bq_0^{\ast}}+  \bp_0 \right)\, . 
\end{equation}
The group $\mathrm{SIM}(2)$ is not unimodular. The left invariant Haar measure with respect to \eqref{leftaction} is given by the Euclidean measure:
\begin{equation}
\ud^2{ \bq }\,\ud^2{ \bp }= \ud q_1\,\ud q_2\,\ud p_1\,\ud p_2\, , 
\end{equation}
which justifies our choice of parametrization \eqref{qpathtp}.
 This is easily  proved from $\ud^2 \bq^{\prime}=q_0^{2}\ud^2 \bq $ and $\ud^2 \bp^{\prime}=\dfrac{\ud^2 \bp }{q^2_0}$. 
 %\textcolor{red}{with both $\ud^2 \bq $ and $\ud^2 \bp $ being rotationally invariant}.  
We also notice that $\dfrac{\ud^2 \bq }{q^2}$ is invariant under complex multiplication and inversion.

\section{An outline of 2D covariant affine integral quantization}
\label{2dcovaffq}
%\subsection{An outline of the formalism}
%\label{genform}
Here we recall the main formulae concerning 2D ACIQ, extracted  from  \cite{gakomu20} and the erratum \cite{gakomu22}. 
The $2$-D affine group SIM$(2)$ has one unitary irreducible representation $U$ which  is realized
in the Hilbert space of square integrable functions on the configuration space,
\begin{equation}
\label{hilberts}
\mathcal{H}=L^{2}(\mathbb{R}^{2}_\ast,\mathrm{d}^2 \bx)\, , 
\end{equation} 
as 
\begin{equation}
(U( \bq , \bp )\psi)( \bx  )=\frac{e^{\ii  \bp \cdot \bx  }}{q}\psi\left(\frac{ \bx  }{ \bq }\right)\,.\label{affrep+}
\end{equation}
Let us choose a function  $\vap( \bq , \bp )$ on the phase space $\Gamma\simeq$ SIM$(2)$ with the following assumptions: 
\begin{assum}
\label{assumv}
\begin{enumerate}
  \item[(i)] The  function $\vap( \bq , \bp )$ is $C^{\infty}$ on $\Gamma$.
  \item[(ii)] It defines a tempered distribution with respect to the variable $ \bp $ for all $ \bq \neq \mathbf{0}$. 
 \item[(iii)]  The operator $\sfMv$  defined as 
 \begin{equation}
\label{affbop}
\sfMv :=\int_{\Gamma}\, \frac{\mathrm{d}^2 \bq \,\mathrm{d}^2 \bp}{4\pi^2}\,\varpi( \bq , \bp )\,QU( \bq , \bp)Q  \, ,\quad Q\psi( \bx  ):= x\psi( \bx )\, , \quad x=\vert \bx \vert\, ,
\end{equation} 
 is  self-adjoint bounded on $\calH$.
\end{enumerate}
\end{assum}
In what follows, we shall refer to this possibly complex-valued function, somewhat abusively, as the ``weight'' function. Note that one   retrieves the function $\vap$ through the following inversion formula:
\begin{equation}
\label{InversionMom}
\vap( \bq , \bp )=\frac{1}{q}\overline{\mathrm{Tr}\left\{U( \bq , \bp )\sfMv\right\}}\, . 
\end{equation}
with $\overline{\mathrm{Tr}\left\{U( \bq , \bp )\sfMv\right\}}$ denoting the complex conjugation of the expression. The following  properties are proved in \cite{gakomu20}.
\begin{enumerate}
  \item[(i)] Any function $\varpi( \bq , \bp )$ defining  the bounded self-adjoint $\sfMv$ obeys the following symmetry:
\begin{equation}
\label{eq:condition1}
\varpi( \bq , \bp )=\frac{1}{q^2}\overline{\varpi\left( \bq^{-1},- \bq^{*} \bp \right)}\,.
\end{equation}
  \item[(ii)] We have the trace formula:
  \begin{equation}
\label{tracecdmU}
 \mathrm{Tr}\left(\cdm^{-1} U( \bq , \bp )\cdm^{-1}\right)= \delta( \bq -1)\delta( \bp )\, ,
\end{equation} 
where $\delta(\bx):= \delta(x_1)\,\delta(x_2)$, and $\cdm$ denotes the (Duflo-Moore \cite{duflo76}) multiplication operator
%\begin{equation}
%\label{CDM}
$\mathsf{C}_{\mathrm{DM}}\psi( \bx  ) := \dfrac{2\pi}{x}\psi( \bx  )\equiv \dfrac{2\pi}{Q}\psi( \bx  )$.
%\end{equation}
  \item[(iii)] Suppose that the operator $\sfMv$ is unit trace class, $\mathrm{Tr}(\sfMv) = 1$. Then  its corresponding   function $\vap( \bq , \bp )$  obeys
\begin{equation}
\label{Tr1cond}
  \vap(1,\mathbf{0})= 1\, . 
\end{equation}
\end{enumerate}
The key result derived from Assumption \ref{assumv}, which is central for the present study, is the resolution of the identity in the Hilbert space $\mathcal{H}$ (see \eqref{hilberts}), which carries the representation defined in \eqref{affrep+}:
\begin{equation}
\label{resolun}
 \int_{\Gamma}\frac{\ud^2  \bq \,\ud^2  \bp }{c_{\sfMv}} \, \sfMv( \bq , \bp ) = \bun_{\mathrm{d}}\,, 
\end{equation}
where the operator-valued function  $\sfMv( \bq , \bp )$ is obtained from a fixed operator $\sfMv$ through the  $U$-transport
\begin{equation}
\sfMv( \bq , \bp )=U( \bq , \bp )\sfMv U^{\dag}( \bq , \bp )\,.  
\end{equation}
The explicit expression of the constant $c_{\sfMv}$ is given in Proposition \ref{propq} below. 
According to  \eqref{resolun}, the quantization of a function (or more generally a distribution) $f(\bq , \bp )$ defined on the phase space $\Gamma\simeq$ SIM$(2)$ is realized through the linear map $f\mapsto \mathrm{Op}^{\vap}_f$   where the operator $\mathrm{Op}^{\vap}_f$  acts in the Hilbert space $\mathcal{H}$: 
\begin{equation}
\label{genqvap}
f\mapsto \mathrm{Op}^{\vap}_f= \int_{\Gamma}\frac{\ud^2  \bq \,\ud^2  \bp }{c_{\sfMv}} \, f( \bq , \bp )\, \sfMv( \bq , \bp )\,, \quad \mathrm{Op}^{\vap}_{f=1}= \bun_{\mathrm{d}}\,.
\end{equation}
The integral on the right hand side of Equation \eqref{genqvap}  is understood as the sesquilinear form, 
\begin{equation}
\label{affquantform}
B_f(\phi_1,\phi_2):= \int_{\Gamma}\frac{\ud^2  \bq \,\ud^2  \bp }{c_{\sfMv}} \, f( \bq , \bp )\, \left\lg \phi_1\left | \sfMv( \bq , \bp ) \right | \phi_2\right\rg,
\end{equation}
for all $\phi_1$, $\phi_2$ in the dense linear subspace $C_0^{\infty}({\R_{*}^2})$. 
When it is defined in this sense, the quantization map \eqref{genqvap} is covariant with respect to the unitary
$2$-D affine action $U$:
\begin{equation}
\label{covaff} U( \bq _0, \bp _0) \hat{\mathrm{A}}^\vap_f U^{\dag}( \bq _0, \bp _0) =
\hat{\mathrm{A}}^\vap_{\mathfrak{U}( \bq _0, \bp _0)f}\, ,
\end{equation}
with
\begin{equation}
\label{covaff2}
 \left(\mathfrak{U}( \bq _0, \bp _0)f\right)( \bq , \bp )=
f\left(( \bq _0, \bp _0)^{-1}( \bq , \bp )\right)= f\left(\frac{ \bq }{ \bq _0},\bq_0^{\ast}( \bp 
- \bp _0) \right)\, ,
\end{equation}
 In view of practical calculations we express  $A^\vap_f$ as an integral operator in $\calH$,  if, of course,  the function or distribution $f$ allows such a characterization. 
\beprop
\label{propq}
Besides the 
{\rm Assumption 1} on the function $\vap( \bq , \bp )$, let us suppose that the function 
\begin{equation}
\label{Omu0}
\Omega( \bq):= \int_{\mathbb{R}_{\ast}^2}\frac{\ud^2  \bx }{x^{2}}\,\widehat{\vap}_p\left( \bq,\pm \bx \right)\, , 
\end{equation}
obeys 
\begin{equation}
\label{condom1}
0< \Omega(1) < \infty\,.
\end{equation}
In \eqref{Omu} 
 $\widehat{\vap}_p$ is the partial $2$-D Fourier transform of $\vap$ with respect to the variable $ \bp $:
\begin{equation}
\label{parcoure} 
\widehat{\vap}_p( \bq , \bx  )= \frac{1}{2\pi}\int_{\mathbb{R}^2} \ud^2 \bp \, e^{-\ii  \bp \cdot \bx  } \vap( \bq , \bp )\,.
\end{equation}
The convergence of the integral in \eqref{Omu0} imposes that  
\begin{equation}
\label{behav0}
\widehat{\vap}_p( \bq , \bm{0})=0\, .
\end{equation}  
Suppose that the function $f$ defines  $ \mathrm{Op}^{\vap}_f$ as a quadratic form  in the sense of \eqref{affquantform}. Then the action of $ \mathrm{Op}^{\vap}_f$ on $\phi$ in $C_0^{\infty}(\R_{*}^{2})$ is given in the form of the linear integral  operator
\begin{equation}
\label{acMom3}
(\mathrm{Op}^{\vap}_f\phi)( \bx  ) = \int_{\R_{\ast}^2}\mathcal{A}^{\vap}_f( \bx  , \bx^{\prime})\,\phi( \bx^{\prime})\,\ud^2   \bx^{\prime}\, ,
\end{equation}
where the kernel $\mathcal{A}^{\vap}_f$ reads as
\begin{equation}
\label{kerMomB}
\mathcal{A}^{\vap}_f( \bx  , \bx^{\prime}) 
  = \frac{1}{c_{\sfMv}}\, \frac{x^2}{{x^{\prime}}^2}\int_{\mathbb{R}_{\ast}^2}\frac{\ud^2  \bq }{q^2}\,\widehat{\vap}_p\left(\frac{ \bx  }{{\bx}^{\prime}},- \bq \right)\, \hat{f}_p\left(\frac{ \bx  }{ \bq }, \bx^{\prime}- \bx  \right)\, , 
\end{equation}
  and the constant $c_{\sfMv}$ is given by
 \begin{equation}
\label{cMom}
c_{\sfMv}= 2\pi \Omega(1)\,. 
\end{equation} 

Here $\hat{f}_p$ is the partial Fourier transform of $f$ with respect to the variable $ \bp $ as is defined in Eq.\;\eqref{parcoure}.
\enprop

\begin{figure}[H]
\begin{center}
\includegraphics[width=6.4in]{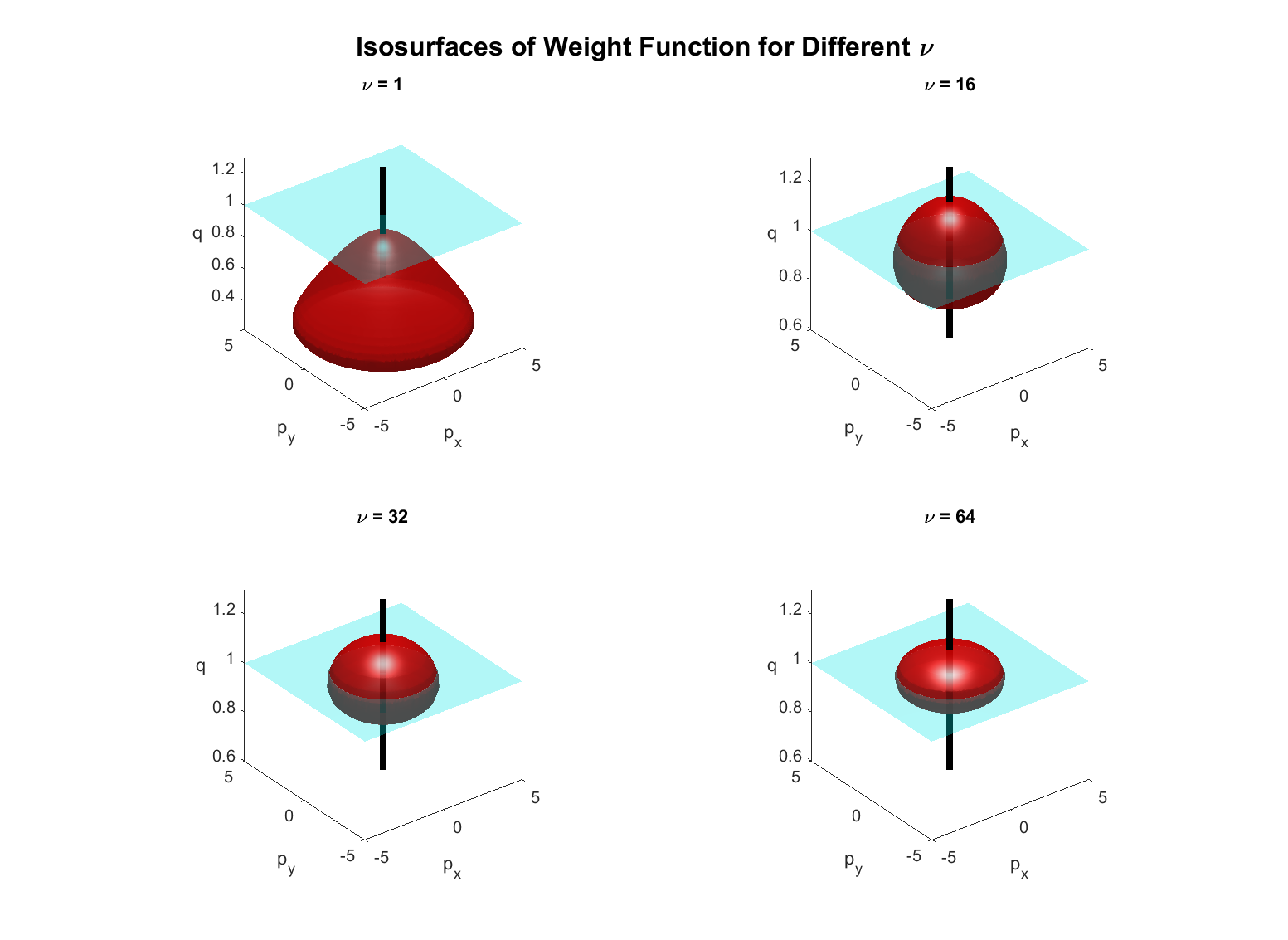}
\caption{Phase-space representation of the quantization weight function $\varpi(\bq,\bp)$ defined in \eqref{exom1}, shown for angle $0\ \text{rad}$ and strictly positive $\bq = q$. The function is plotted over the variables $p_1$, $p_2$, and $q$, with the constant $\sigma$ fixed at $3.5$, and for four different values of the parameter $\nu$ (as indicated in the figure). As $\nu$ increases, the weight function becomes increasingly localized around the  point $(\bq = (1,0), \bp = (0,0))$ in phase space, i.e., the unit of the group SIM$(2)$. For visualization purposes, the function has been normalized by its maximum modulus, and the isovalue contour level is set at $0.5$.
}
\label{isosur}
\end{center}
\end{figure}

In the following, we also consider a broader class of functions, of which \eqref{Omu0} represents a particular case:
\begin{equation}
\label{Omu}
\Omega_\beta( \bq):= \int_{\mathbb{R}_{\ast}^2}\frac{\ud^2  \by }{y^{\beta +2}}\,\widehat{\vap}_p\left( \bq,\pm \by \right)\, , \quad \Omega_0 ( \bq)\equiv  \Omega( \bq)\, ,
\end{equation}
with suitable conditions on $\varpi$ for having convergence of this integral in the neighborhood of $\bm{0}$.
Hence,  the condition \eqref{Tr1cond} can be also written as a condition on the function $\Omega_{-2}$.
\begin{equation}
\label{Tr2cond}
\mathrm{Tr}(\sfMv)= \frac{1}{2\pi}\int_{\mathbb{R}_{\ast}^2}\ud^2  \bx   \,\widehat{\vap}_p(1,- \bx  )=  \frac{1}{2\pi} \Omega_{-2}(1)=  1\, .
\end{equation}
Let us give the following example of a function $\varpi$ which complies with conditions \eqref{InversionMom}, \eqref{Tr1cond}, \eqref{condom1}, and \eqref{behav0}. It is a function well-localised about the affine identity $(1,\bm{0})$ in $\Gamma$:
\begin{equation}
\label{exom1}
\vap( \bq , \bp )= \frac{e^{2\nu}}{q}\, e^{-\nu\left( q + 1/q\right)}\, \alpha(\arg(\bq))\,\left(1- q\dfrac{p^2}{2\sigma^2}\right)\,e^{-q\frac{p^2}{2\sigma^2}}\, , \quad \arg(\bq)= \arctan\frac{q_2}{q_1}\, ,
\end{equation}
where $\nu$ and $\sigma$ are two positive parameters, and the (possibly) complex-valued smooth and bounded  function $\alpha$ obeys:
\begin{equation}
\label{condalpha}
\overline{\alpha(\arg(\bq))}= \alpha(-\arg(\bq))\, , \quad \mbox{and} \quad \alpha(\bm{0})=1\,. 
\end{equation}
A simple example of $\alpha$ is 
\begin{equation}
\label{exalpha}
\alpha(\arg(\bq))= \exp\ii\mu\arg(\bq)\,, 
\end{equation}
where $\mu$  is a real constant. 
In Figure~\ref{isosur}, we display the phase-space representation of the quantization weight function $\varpi(\bq,\bp)$ defined in \eqref{exom1} for several values of $\nu$ (namely, $1, 16, 32$ and $64$), fixing the parameter $\sigma = 3.5$. As $\nu$ increases, the weight function becomes progressively more localized around the point $(\bq = (1,0), \bp = (0,0))$, which corresponds to the identity element of the group SIM$(2)$, i.e, the origin of the affine phase space.

The corresponding function $\Omega(\bq)$ is given by:
\begin{equation}
\label{Omex1}
\Omega(\bq)= \frac{\pi \sigma^2 e^{2\nu}}{q^2}\, e^{-\nu\left( q + 1/q\right)}\, \alpha(\arg(\bq))\, , \quad \Omega(1)= \pi\sigma^2\,. 
\end{equation}
Implementing the action \eqref{acMom3} of the operator $\mathrm{Op}^{\vap}_f$ for specific choices of $f$ is generally a nontrivial task, as illustrated in Appendices \ref{app:examples=quantization} and \ref{rankone}, where we provide methodological details and work through several representative examples. We focus here on the cases most pertinent to the aims of the present study.

\begin{enumerate}
\item  $f( \bq , \bp )= \bp $\\ 
\begin{equation}
\label{qmom}
\mathrm{Op}^{\vap}_{ \bp } =  \bP  + \frac{\ii}{\bQ^{\ast}}\left(2+\frac{\pmb{\nabla}\Omega(1)}{\Omega(1)}\right)\, . 
\end{equation}
\item  $f( \bq , \bp )= \bp  ^{\,2}$\\ 
 \begin{equation}
 \label{qkin}
  \mathrm{Op}^{\vap}_{ \bp  ^{\,2}}= \bP^2+\frac{2\ii}{\bQ^{\ast}}\left(2+\frac{\pmb{\nabla}\Omega(1)}{\Omega(1)}\right)\cdot  \bP   -\frac{1}{Q^2}\left[4+4\frac{\mathbf{e}_1\cdot\pmb{\nabla}\Omega(1)}{\Omega(1)}+\frac{\triangle\Omega(1)}{\Omega(1)}\right]\,,
 \end{equation}
 where $\pmb{\nabla}\Omega(1)$ and $\triangle\Omega(1)$ are  given by
\begin{equation*}
\label{ }
\pmb{\nabla}\Omega(1)= \int_{\mathbb{R}_{\ast}^2}\frac{\ud^2  \bx}{x^{2}}\,\left.\pmb{\nabla}_{\bq}\widehat{\vap}_p\left( \bq, \bx \right)\right\vert_{\bq=1}\,, \quad  \triangle\Omega(1)= \int_{\mathbb{R}_{\ast}^2}\frac{\ud^2  \bx}{x^{2}}\,\left.\triangle_{\bq}\widehat{\vap}_p\left( \bq, \bx \right)\right\vert_{\bq=1}\,.
\end{equation*}

 \end{enumerate}

\section{Emerging vector and scalar fields from affine quantization}
\label{geompot}
We now examine the physical content of  the expressions \eqref{qmom} and \eqref{qkin} for the quantum momentum  and kinetic energy respectively
for the free particle in the punctured plane. 
In order to ensure that the action remains invariant under local gauge transformations,
a gauge transformation of the wave function, $\psi(t,\mathbf{x}) \mapsto e^{\mathrm{i}\Lambda(t,\mathbf{x})}\psi(t,\mathbf{x})$,
must be accompanied by a corresponding transformation of the gauge fields. Consequently, the Schr\"{o}dinger equation naturally incorporates the scalar and vector potential operators via the minimal coupling procedure \cite{jmll67,jmll74}.
We are thus prompted by \eqref{qmom} to introduce the concept of the \textit{affine vector potential}:
\begin{equation}
\label{vecpot}
\bA^{\vap}(\bQ):= 
- \frac{\ii}{\bQ^{\ast}}\left(2+\frac{\pmb{\nabla}\Omega(1)}{\Omega(1)}\right)\,.  
\end{equation}
Let us rewrite the affine quantum kinetic operator \eqref{qkin} (up to a factor like $1/2m$) in the more explicit form: 
\begin{align}
\label{qp2} 
  \mathrm{Op}^{\vap}_{ \bp  ^{\,2}}
&= \bP^2+\frac{2\ii}{\bQ^{\ast}}\left(2+\frac{\pmb{\nabla}\Omega(1)}{\Omega(1)}\right)\cdot  \bP   -\frac{1}{Q^2}\left[4+4\frac{\mathbf{e}_1\cdot\pmb{\nabla}\Omega(1)}{\Omega(1)}+\frac{\triangle\Omega(1)}{\Omega(1)}\right] \\\nonumber
&={\left[{\bf P} -\ii 
{\begin{pmatrix}
-2 - \frac{\partial_{1} \Omega (1)}{\Omega (1)} &\frac{\partial_{2} \Omega (1)}{\Omega (1)}  \\       
- \frac{\partial_{2} \Omega(1)}{\Omega (1)} &- 2 - \frac{\partial_{1} \Omega (1)}{\Omega (1)}  \\    
\end{pmatrix}} \frac{\bQ}{Q^2}\right] }^2\!\! 
  +\frac{\pmb{\nabla}\Omega(1)\cdot\pmb{\nabla}\Omega(1)- \Omega(1)\triangle\Omega(1)}{(\Omega(1))^2}\frac{1}{Q^2}\\
&=  \left(\bP - \mathfrak{q}\,\bA^{\vap}(\bQ) \right)^2 +\frac{ K^{\vap}}{Q^2}  \, ,
\end{align}
where $[Q_i, P_j] = \ii \bun_{\mathrm{d}}$, $\mathfrak{q}$ is a ``fictive'' charge, and
\begin{align}
    \bA^{\vap}(\bQ) 
    &\equiv \frac{\ii}{\mathfrak{q}} \,
{\begin{pmatrix}
-2 - \frac{\partial_{1} \Omega (1)}{\Omega (1)} &\frac{\partial_{2} \Omega (1)}{\Omega (1)}  \\       
- \frac{\partial_{2} \Omega(1)}{\Omega (1)} &- 2 - \frac{\partial_{1} \Omega (1)}{\Omega (1)}  \\    
\end{pmatrix}} \frac{\bQ}{Q^2} \, ,\\
K^{\vap} &\equiv \frac{\pmb{\nabla}\Omega(1)\cdot\pmb{\nabla}\Omega(1)- \Omega(1)\triangle\Omega(1)}{(\Omega(1))^2} \, .
\end{align}
In this derivation, we utilized formulae defined in Appendix \ref{app:formulae}, and 
chose the weight $\vap(\bq,\bp)$ in Eq.\ (\ref{Omu}) satisfying  
\begin{eqnarray}
\partial_{1} \ln \Omega(1) = - 2  \label{eqn:partial_1omega} \, .
\end{eqnarray}
Then the affine vector potential is simplified as
\begin{equation}
 \bA^{\vap}(\bQ) = A^\vap_{x_1}(\bQ) {\bf e}_1 + A^\vap_{x_2}(\bQ) {\bf e}_2 \, ,
\end{equation}
where
\begin{equation}
A^\vap_{x_1}(\bQ) = - \frac{1}{2\pi } \Phi^\vap_0 \frac{Q_2}{Q^2} \, , \qquad A^\vap_{x_2}(\bQ)
 = \frac{1}{2\pi } \Phi^\vap_0 \frac{Q_1}{Q^2} \, ,
\end{equation}
with the topological ``affine'' magnetic flux, 
\begin{eqnarray}
\Phi^\vap_0 = - \ii  \frac{2\pi}{\mathfrak{q}}\partial_2 \ln \Omega (1)  \, , \label{eqn:partial_2omega}
\end{eqnarray}
or, after restoring physical dimensions,
\begin{equation}
\Phi^\vap_0 = - \ii  \frac{2\pi\hbar}{\mathfrak{q}}\partial_2 \ln \Omega (1)  \, , \label{eqn:partial_3omega}
\end{equation}
 thereby making explicit the appearance of the celebrated Dirac magnetic flux quantum $h/\mathfrak{q}$. 
With the choice \eqref{condalpha}, leading to the function $\Omega$ in \eqref{Omex1}, the condition \eqref{eqn:partial_1omega} is satisfied, and the resulting flux is given by:
\begin{equation}
\label{flux2}
\Phi^\vap_0 = - \ii  \frac{2\pi\hbar}{\mathfrak{q}}\,\alpha^{\prime}(0)  \, .
\end{equation}
With the more specific choice \eqref{exalpha} the flux takes the form 
\begin{equation}
\label{flux3}
\Phi^\vap_0 =  \frac{2\pi\hbar \mu}{\mathfrak{q}} \, .
\end{equation}
Thus, by merely respecting the symmetry structure of the model, and  if we choose $\mu=n\in \Z$, i.e., if the chosen weight function $\vap$ is $2\pi$-periodic in its $\bq$ variable, our affine vector potential, equivalently the derived magnetic flux,  become discretized. This feature is analogous, though distinct, from the phenomenon of flux quantization due to the single-valuedness of the wave function, as observed when an electron orbits the singularity at the origin.
Comparing these with Eq.\ \eqref{vecfield}, one can see that the affine vector potential is the same as the AB gauge potential.
A similar vector potential is obtained in the modified canonical quantization on a 1D loop, as is shown by Equation (3.6) in \cite{ohnukisis95}.

In the system under consideration, there exists a domain where the wave function cannot occupy the origin of the coordinates. 
This phenomenon arises due to the last term on the right-hand side of Eq. (\ref{qp2}), which is referred to as the {\it affine scalar potential}. 
This term does not appear in \cite{ohnukisis95}, and when $K^{\vap}$ is positive, it produces a repulsive, centrifugal-like effect. In a semi-classical interpretation, this additional centrifugal term prevents the electron from reaching the singular point, regardless of its angular momentum \cite{gazkoimur17}. 
Moreover, by selecting an appropriate $\varpi$, this coefficient can be made sufficiently large, ensuring that the operator (\ref{qp2}) becomes essentially self-adjoint \cite{reedsimon2,kowal02,asch07}. 
For instance, with the choice \eqref{condalpha}, we obtain (with physical units):
\begin{equation}
\label{Kvapex}
K^{\vap}= \hbar^2 \left(2\nu + (\alpha^{\prime}(0))^2 - \alpha^{\prime\prime}(0)\right)\,.
\end{equation}
With the choice \eqref{exalpha} this reads 
\begin{equation}
\label{Kvapex1}
K^{\vap}= 2 \hbar^2 \nu^2\,.
\end{equation}
Thanks to the presence of the arbitrary parameter $\nu > 0$, this quantity can be made sufficiently large to ensure that the quantum Hamiltonian \eqref{qp2} is self-adjoint. 
In other words, while conventional canonical quantization, based on Weyl-Heisenberg symmetry, introduces quantum-level ambiguities (as was discussed in the introduction),  
affine covariant integral quantization removes these ambiguities by employing a suitable choice of $\varpi$.

\section{Conclusion}
\label{disc}

In this study, we have explored the quantum dynamics of a particle in a punctured plane using 2-D affine covariant integral quantization (ACIQ). The phase space is identified with the similitude group SIM(2) in two dimensions, allowing for covariant integral quantization. 
A preliminary, comprehensive and accessible,  exposition of the mathematical material was indispensable to comprehending the genuinely novel contribution of this paper: the reinterpretation of the Aharonov-Bohm (AB) effect as an emergent consequence of the symmetry structure of the model at the quantum level, as revealed through the ACIQ framework.

Our ACIQ quantization scheme  preserves the geometric structure of the phase space and its affine symmetry and reveals both vector and scalar potentials. Upon restoring physical dimensions, these expressions are found to be proportional to the Planck constant, highlighting a purely quantum effect. This is the central outcome of our work.
More precisely, around the punctured point, we found that the topology that influences quantum fluctuations induces the affine vector potential, which can be identified with the AB gauge potential formed by the infinite coil.
This suggests that the AB effect arises from the topology imposed by the impenetrable coil, rather than the classical gauge potential. Naturally, once the potentials are obtained, the Aharonov-Bohm effect can also be derived through a suitable choice of the affine quantizer function, see \eqref{exom1} and \eqref{exalpha}. Our findings provide a novel perspective on the AB effect, emphasizing the role of topology in quantum mechanics. 

In ACIQ, however, topology induces not only the vector potential but also an affine scalar potential, which 
prevents an electron from reaching the singular point and is inevitable to satisfy essentially self-adjointness. 
The scalar potential introduced by ACIQ modifies the electron's trajectory near the coil, creating a barrier effect that prevents the standard AB interference. Future studies should quantify this deviation to clarify its implications for the AB effect.

The results suggest potential extensions to three-dimensional systems, such as monopoles in ${\R}^3$, where the configuration space is ${\R}_{*}^{3}$ \cite{Cardona, Carinena,lyndnou98} and the phase space is ${\R}_{*}^{3}\times\R^3$. This extension could involve identifying the phase space with a left coset of the similitude group in three dimensions by SO$(2)$ or the cotangent bundle on ${\R}_{*}^{3}$. These generalizations highlight the broader implications of the study, suggesting that the methods and results could be applied to a wide range of physical phenomena beyond the specific case of the Aharonov-Bohm effect.

Certainly, the quantization method in more general spaces with non-trivial topology remains to be fully developed. Further investigation into the relationship between the Aharonov-Bohm effect and quantum topological phenomena is necessary. The challenge lies in the fact that, beyond the specific scenario addressed in this work, there is generally no readily identifiable and applicable symmetry.

\section*{Acknowledgments}
T.K. acknowledges the financial support by CNPq (No.\ 305654/2021-7) 
and the Fueck-Stiftung.
A part of this work has been done under the project INCT-Nuclear Physics
and Applications (No.\ 464898/2014-5). R. Murenzi and A. Zlotak are grateful for the funding provided by
the Physics Department of the Worcester Polytechnic Institute. J.-P. Gazeau gratefully acknowledges the financial support provided by Worcester Polytechnic Institute, which facilitated his participation in the ongoing collaboration underlying this work.

\appendix
\section{Some observations related to canonical quantization } \label{app:canonical}

The discussion in this section is the summary of the discussion given in \cite{book_ohnuki}.

When we introduce a velocity operator instead of the momentum operator, we can eliminate the appearance of the gauge potentials in the operator equations of quantum mechanics. In this case, the commutation relations are defined by 
\begin{align}
[Q_j,Q_k] &= 0 \, , \\
[Q_j, \dot{Q}_k] &= \ii \frac{\hbar}{m} \delta_{jk} \mathbbm{1}\, ,\\
[\dot{Q}_j, \dot{Q}_k] &= \frac{\ii \mathfrak{q} \hbar}{m^2} \sum_{l=1,2,3} \epsilon_{jkl} B_l ({\bf Q}) \, ,
\end{align}
where $Q_i\psi(\bx)= x_i\psi(\bx), i=1,2,3$. 
In this case, because the Hamiltonian operator is independent of the gauge potential,  
the effect of the gauge potential does not appear in any operator equation.
Then why do we observe the Aharonov-Bohm effect in this case?

To understand this, we have to remember that the commutation relations are not enough to study quantum dynamics and we need to investigate 
the representations of the self-adjoint operators satisfying these relations.
Suppose that the eigenvectors of the position operator form a complete set.
Then from the second equation, we find
\begin{eqnarray}
(x_j - x^\prime_j) \langle {\bf x} | \dot{Q}_k | {\bf x}^\prime \rangle = \ii \frac{\hbar}{m} \delta_{jk} \delta^{(3)}({\bf x} - {\bf x}^\prime) \, .
\end{eqnarray}
The general solution to satisfy this equation is 
\begin{eqnarray}
\langle {\bf x} | \dot{Q}_k | {\bf x}^\prime \rangle= -\frac{\hbar}{m} \left\{ 
\ii \frac{\partial}{\partial x_k} + C_k ({\bf x}) 
\right\} \delta^{(3)}({\bf x} - {\bf x}^\prime) \, ,
\end{eqnarray}
where $C_k ({\bf x})$ is an arbitrary function of ${\bf x}$.

Next, from the third equation, we find
\begin{align}
&\langle {\bf x} | [\dot{Q}_j, \dot{Q}_k] |{\bf x}^\prime \rangle 
=
\int \ud^3 {\bf x}^{\prime\prime} 
\left\{
\langle {\bf x} |\dot{Q}_j |{\bf x}^{\prime\prime} \rangle 
\langle {\bf x}^{\prime\prime} |\dot{Q}_k |{\bf x}^{\prime} \rangle 
-
(j \leftrightarrow k)
\right\} \nonumber \\
& =
 \ii\frac{\hbar^2}{m^2} \left\{
\frac{\partial C_k ({\bf x})}{\partial x_j} - \frac{\partial C_j ({\bf x})}{\partial x_k}
\right\} \delta^{(3)}({\bf x} - {\bf x}^\prime)
= 
\frac{\ii \mathfrak{q} \hbar}{m^2} \sum_{l=1,2,3} \epsilon_{jkl} B_l (\bf Q) \delta^{(3)}({\bf x} - {\bf x}^\prime) \, .
\end{align}
If there is no magnetic field, we can set ${\bf C} ({\bf x}) =0$ leading to the standard result in quantum mechanics.
However, when ${\bf C} ({\bf x}) \neq 0$, we find 
\begin{eqnarray}
\nabla \times {\bf C} ({\bf x}) = \frac{\mathfrak{q}}{\hbar} {\bf B} ({\bf x}) \, .
\end{eqnarray}
That is, the function ${\bf C} ({\bf x})$ is nothing but the gauge potential.

In short, to construct self-adjoint operators satisfying the above commutation relations, 
we need to introduce a degree of freedom equivalent to the gauge potential.

\section{Quantization examples}
\label{app:examples=quantization}

\subsection{Separable functions:} \begin{center}
$f( \bq , \bp )\equiv u( \bq )\,v( \bp )$
\end{center}
 
 The formula \eqref{kerMomB} simplifies to 
\begin{equation}
\label{PosMomA}
\mathcal{A}^{\vap}_{u( \bq )v( \bp )}( \bx  , \bx^{\prime})=  \frac{1}{c_{\sfMv}}\,\hat v( \bx^{\prime}- \bx  )\, \frac{x^2}{{x^{\prime}}^2}\, \left(\widehat{\vap}_p\left(\frac{ \bx  }{ \bx^{\prime}},-.\right)\,\ast_{aff} u\right)( \bx )\,, 
\end{equation}
where 
\begin{itemize}
  \item the \textit{$2$D-affine convolution product} $\ast_{aff} $ on the multiplicative group $\R^{2}_{*}=\R^{+}_{*}\times \mbox{SO}(2)$ is defined by
\begin{equation}
\label{conv2d}
(f_1\ast_{aff} f_2)( \bx  )=\int_{\R_{\ast}^2}\frac{\ud^2 \bx^{\prime}}{{x^{\prime}}^2}\,f_1( \bx^{\prime})f_2\left({\frac{ \bx  }{ \bx^{\prime}}}\right)\,,
\end{equation}
\item the dot inside the r.h.s. parenthesis in \eqref{PosMomA} stands for the integration variable in the affine convolution product,
  \item  $\hat v$ is the Fourier transform of $v$.
\end{itemize}

\subsection{Position dependent functions }
\begin{center}
$f( \bq , \bp )\equiv u( \bq )$
\end{center}
 The expression \eqref{PosMomA} then yields the multiplication operator
\begin{equation}
\label{quq}
\left(\mathrm{Op}^{\vap}_{u( \bq )}\phi\right)( \bx  )=\frac{1}{c_{\sfMv}}(w*_{aff}u)( \bx  ) \phi( \bx  )\, , 
\end{equation}
where $w( \bx  )=2\pi\widehat{\vap}_p(1,- \bx )$.
Note the particular cases: 
\begin{enumerate}
\item $u( \bq )$ is a simple power of $q$, say $u( \bq )= q^{\beta}$. Then we have \begin{equation}
\label{Aqbeta}
\mathrm{Op}^{\vap}_{q^{\beta}}=  \frac{2\pi}{c_{\sfMv}}\int_{\R_{\ast}^2}\frac{\ud^2 \by}{y^{2+\beta}}\,\widehat{\vap}_p(1,-\by)\,Q^{\beta}=  \frac{\Omega_{\beta}(1)}{\Omega(1)}\, Q^{\beta}\, . 
\end{equation}
\item  $u( \bq )= \bq $ stands for the  classical vector position in the plane, 
%$\Omega_{(2,0,1)}(1)=0$, \red{\textbf{WHY? PLEASE CHECK! SEE ALSO  BELOW WHERE WE SHOULD THEN HAVE A DIAGONAL MATRIX !}}
\begin{equation}
\label{Aq1-2}
\mathrm{Op}^{\vap}_{ \bq }
=  \frac{2\pi}{c_{\sfMv}}{\begin{pmatrix}
 \Omega_{(2,1,0)}(1) & \Omega_{(2,0,1)}(1) \\       
 -\Omega_{(2,0,1)}(1) & \Omega_{(2,1,0)}(1) \\      
\end{pmatrix}}
{\begin{pmatrix}
Q_1 \\       
Q_2 \\      
\end{pmatrix}}\, , 
\end{equation} 
where we have generalized the notation \eqref{Omu} for the following integrals
\begin{equation}
\label{Omuqq}
\Omega_{\beta, \nu_1,\nu_2}( \bu):= \int_{\mathbb{R}_{\ast}^2}\frac{\ud^2 \by}{y^{\beta +2}}\,\widehat{\vap}_p\left( \bu,-\by\right)\,y_1^{\nu_1}\,y_2^{\nu_2}\,, \quad \Omega_{\beta,0,0}( \bu)\equiv \Omega_{\beta}( \bu)\,,
\end{equation}
with $\by= \begin{pmatrix}
      y_1    \\
      y_2  
\end{pmatrix}$\,. 
\end{enumerate}

For the special case of 
\begin{align*}
    \Omega_{(2,0,1)}(1) &= 0\\
    \Omega_{(2,1,0)}(1) &= \frac{c_{M^\varpi}}{2\pi},
\end{align*}
our matrix is the unit matrix $\mathbbm{1}_2$. This results in the operator $\mathrm{Op}^{\vap}_{ \bq }$ being a multiplication operator via $Q$. One such example that satisfies this condition is
\begin{align}
    \hat{\varpi}_p(1,-\bq) = q^4 e^{-\frac{1}{2}q^2}q_1.
\end{align}

\subsection{Monomial functions in momentum coordinates  }

\begin{center}
$f( \bq , \bp )\equiv u( \bq )\, p_i^n$
\end{center}
\begin{equation}
\mathcal{A}^{\vap}_{u( \bq )\, p_i^n}( \bx  , \bx^{\prime})=\frac{{\ii}^n\;2\pi}{c_{c_{\sfMv}}}\frac{x^2}{{x^{\prime}}^2}
\delta({{x}^{\prime}}_{3-i}-x_{3-i})\,\delta^{(n_i)}_{x^{\prime}_i}({x^{\prime}}_i-x_i)
\, \frac{x^2}{{x^{\prime}}^2}\, \left(\widehat{\vap}_p\left(\frac{ \bx  }{ \bx^{\prime}},-.\right)\,\ast_{aff}u\right)(\bx)\, . 
\end{equation}
Applying the corresponding operator on $\phi \in C^{\infty}\left(\R^2_\ast\right)$ yields the expansion in powers of the operator $P_i = -\ii\partial/\partial x_i$:
 \begin{equation}
 \label{Aupin}
 \left(\mathrm{Op}^{\vap}_{u( \bq )p_{i}^n}\phi\right)( \bx  )= \frac{2\pi Q^2}{c_{\sfMv}}\sum_{s=0}^{n}\binom{n}{s}(-\ii)^{n-s}\frac{\partial^{n-s}}{\partial {{x^{\prime}}_i}^{n-s}}
 \left[\frac{1}{{x^{\prime}}^2}\left(\widehat{\vap}_p\left(\frac{ \bQ }{ \bx^{\prime}},-.\right)\,\ast_{aff}u\right)(\bQ)\right]_{ \bx^{\prime}= \bQ }\,P_i^{s}\phi( \bx  )\,, 
 \end{equation}
 where $\binom{n}{s}$ is the binomial coefficient.
For the lower powers $n$, this formula particularizes as follows. 
\begin{enumerate}
\item  $f( \bq , \bp )=u( \bq )  \bp $
\begin{equation}
%\label{Aubp}
\begin{split}
\mathrm{Op}^{\vap}_{u( \bq ) \bp } &= \frac{2\pi}{c_{\sfMv}}(\widehat{\vap}_p(1,-\cdot)\ast_{aff}u)( \bQ )\, \bP   +\\
&+ \frac{4\pi \ii}{c_{\sfMv}}\frac{1}{\bQ^{\ast}} (\widehat{\vap}_p(1,-\cdot)\ast_{aff}u)( \bQ )
+ \frac{2\pi \ii}{c_{\sfMv}}\frac{1}{\bQ^{\ast}} ((\pmb{\nabla} \widehat{\vap}_p)(1,-\cdot)\ast_{aff}u)( \bQ)\, .
\end{split} 
\end{equation}
Here $(\pmb{\nabla} \widehat{\vap}_p)(1,-\cdot):= \left[\left(\dfrac{\partial}{\partial q_1}\widehat{\vap}_p)(\bq,-\cdot),\dfrac{\partial}{\partial q_2}\widehat{\vap}_p)(\bq,-\cdot\right)\right]_{ \bq= \mathbf{1} }$. More generally and from now on, the  gradient vector $ \pmb{\nabla}F(\bq)$ of a  function $F(\bq)$,  is as well  viewed as the complex number   $\left(\dfrac{\partial}{\partial q_1}F(\bq),\dfrac{\partial}{\partial q_2}F(\bq)\right)$
 associated with $\{\mathbf{e}_1, \mathbf{e}_2\}$.

\item  $f( \bq , \bp )= \bp $\\ Then the previous one simplifies to the quantum momentum:
\begin{equation}
%\label{qmom}
\mathrm{Op}^{\vap}_{ \bp } =  \bP  + \frac{\ii}{\bQ^{\ast}}\left(2+\frac{\pmb{\nabla}\Omega(1)}{\Omega(1)}\right)\, . 
\end{equation}
\item  $f( \bq , \bp )= \bp  ^{\,2}$\\ Eq. \eqref{Aupin} yields the quantum kinetic energy (up to the usual factor $1/2m$):
 \begin{equation}
% \label{qkin}
  \mathrm{Op}^{\vap}_{ \bp  ^{\,2}}= \bP^2+\frac{2\ii}{\bQ^{\ast}}\left(2+\frac{\pmb{\nabla}\Omega(1)}{\Omega(1)}\right)\cdot  \bP   -\frac{1}{Q^2}\left[4+4\frac{\mathbf{e}_1\cdot\pmb{\nabla}\Omega(1)}{\Omega(1)}+\frac{\triangle\Omega(1)}{\Omega(1)}\right]\,,
 \end{equation}
 
 \item $f( \bq , \bp )= \bq \cdot  \bp = q_{1} p_{1}+q_{2} p_{2}$ (generator of dilations in the plane)\\
  By using\eqref{Omuqq}, we get:
\begin{equation}
\label{qdp}
\begin{split}
\mathrm{Op}^{\vap}_{ \bq \cdot  \bp }=&\frac{2\pi}{c_{\sfMv}}\left({\Omega}_{(2,1,0)}(1)( \bQ \cdot \bP   +2\ii)-\Omega_{(2,0,1)}  \bQ \wedge \bP \right.\\&\left. +\ii(\mathbf{e}_1\cdot\pmb{\nabla}\Omega_{(2,1,0)}(1)-\mathbf{ e}_2\cdot\pmb{\nabla}\Omega_{(2,0,1)}(1)) \right)\,.
\end{split}
\end{equation}
\item  $f( \bq , \bp )= \bq \wedge \bp = q_1p_2-q_2p_1$ (angular momentum)\\
 The result is given  in terms of the integrals \eqref{Omuqq}  as
\begin{equation}
\label{qtp}
\begin{split}
\mathrm{Op}^{\vap}_{ \bq \times  \bp }= &
\frac{2\pi}{c_{\sfMv}}\left(\Omega_{(2,1,0)}(1) \bQ \wedge  \bP  +\Omega_{(2,0,1)}(1)( \bQ \cdot \bP  +2\ii)\right.
\\ &\left.+ \ii(\mathbf{ e}_1\cdot\pmb{\nabla}\Omega_{(2,0,1)}(1)+\mathbf{ e}_2\cdot\pmb{\nabla}\Omega_{(2,1,0)}(1))\right)\,.
\end{split}
\end{equation}
 \end{enumerate}
 Simplified expressions for the above formulas are always possible  with a suitable choice of the  function $\vap$. 

\section{Quantization with rank-one density operator}
\label{rankone}

Let us now examine the specific case of rank-one density operator or
projector $\sfM^{\vap_{\psi}}=|\psi\rangle\left\langle \psi\right|$ where $\psi$
is a unit-norm function in
$L^{2}({\R_{\ast}^2},\mathrm{d^2} \bx  ) \cap L^{2}\left({\R_{\ast}^2},\dfrac{\mathrm{d^2} \bx  }{x^2}\right)$. The corresponding function $\varpi\equiv \varpi_{\psi}$ reads as 
\begin{equation}
\label{weightcs}
\varpi_{\psi}( \bq , \bp )=\frac{\langle  U( \bq , \bp ) \psi |\psi \rangle}{q}\, ,  
\end{equation}
and its  partial Fourier transform  $\vap$ is given  by
\begin{equation}
\label{acsvap}
\widehat{\vap}_{\psi}( \bu,\bv)= 2\pi\, \frac{1}{u^2}\, \psi(-\bv)\,\overline{\psi\left(-\frac{\bv}{ \bu}\right)}\,.
\end{equation}
It follows that
\begin{equation}
\label{Omdbeta}
\Omega( \bu)=\frac{2\pi}{u^2}  \int_{\R_{\ast}^2}\frac{\ud^2  \bq }{q^2}\, \psi( \bq )\, \overline{\psi\left(\frac{ \bq }{ \bu}\right)}\,, \quad \Omega(1)=2\pi\left\langle \frac{1}{Q^2}\psi|\psi \right\rangle\equiv 2\pi\left\langle Q^{-2} \right\rangle_{\psi}\,,
\end{equation}
where $\left\langle A \right\rangle_{\psi}$ is the mean value of operator $A$ in the state $|\psi\rg$.
With implicit assumptions on the existence of derivatives of $\psi$ and of their (square)  integrability, we have 
\begin{equation}
\label{Ompsi1a} 
(\pmb{\nabla} \Om)(1)= -2\Omega(1)\mathbf{e}_1-\ii 2\pi\left\langle \frac{ 1 }{ \bQ }\bP\psi|\psi \right\rangle=  -4\pi\left\langle Q^{-2}\right\rangle_{\psi}\mathbf{e}_1 -\ii 2\pi\left\langle \bQ^{-1}\bP\right\rangle_{\psi}\, .
\end{equation}
\begin{equation}
\label{Ompsi1b} 
(\Delta \Omega)(1))=\Omega(1)+8\ii\pi \left\langle \mathbf{ e}_1\cdot\left(\frac{1}{ \bQ } \bP\right)\psi|\psi\right\rangle-2\pi\left\langle  \bP  ^{2}\psi|\psi\right\rangle= 8\pi \left\langle Q^{-2} \right\rangle_{\psi} +8\ii\pi \left\langle Q^{-2}\bQ\cdot \bP\right\rangle_{\psi} -2\pi\left\langle  \bP  ^{2}\right\rangle_{\psi}\, .
\end{equation} 
Therefore, affine vector and scalar potentials particularize as
\begin{equation}
\label{vecpotpsi}
\bA^{\vap_\psi}(\bQ)= - \frac{2\pi}{\Omega(1)} \frac{\bQ}{Q^2}\left\langle \frac{ 1 }{ \bQ }\bP\psi|\psi \right\rangle=  - \frac{Q^{-2} }{\left\langle Q^{-2} \right\rangle_{\psi} }\bQ \left\langle \bQ^{-1}\bP\right\rangle_{\psi}\,. 
\end{equation}
\begin{equation}
\label{scpotpsi}
\mathrm{V}^{\vap_\psi}(\bQ)= \left[\frac{\left\langle  \bP  ^{2}\right\rangle_{\psi}}{\left\langle Q^{-2} \right\rangle_{\psi} } - \frac{\left\langle \bQ^{-1}\bP\right\rangle_{\psi}\cdot\left\langle \bQ^{-1}\bP\right\rangle_{\psi}}{\left\langle Q^{-2} \right\rangle^2_{\psi}}\right]\frac{1}{Q^2}\equiv \frac{K^{\vap_\psi}}{Q^2}\,.
\end{equation}

Note that for $\psi$ real we have:
\begin{equation}\label{psireal1} 
\frac{(\pmb{\nabla} \Om)(1)}{\Omega(1)}=- 2\mathbf{ e}_1 \quad \mbox{i.e.} \quad \left\langle \bQ^{-1}\bP\right\rangle_{\psi}= \mathbf{0}\, , 
\end{equation}
and the above expressions simplify to $\bA^{\vap_\psi}(\bQ)= \mathbf{ 0}$ (no vector potential) and $\mathrm{V}^{\vap_\psi}(\bQ)= \dfrac{\left\langle  \bP  ^{2}\right\rangle_{\psi}}{\left\langle Q^{-2} \right\rangle_{\psi} }\dfrac{1}{Q^2}$.

To make the link with the example \eqref{Omex1}, we now consider  the case where $\psi$ is complex-valued.  One can choose, for example, a real amplitude $g(\bx)$ times a phase function $e^{\mathrm{i}\mu \arg(\bx)}$
that is, $\psi(\bx)= e^{\mathrm{i}\mu \arg(\bx)}g(\bx)$. One obtains: 
\begin{align}
\nonumber\varpi_{\psi}(\bq,\bp)&=\lg U(\bq,\bp)\psi|\psi\rg=\int_{\R^2_{*}}\ud^{2}\bx\,e^{\mathrm{i}\bp\cdot\bx} \,\frac{1}{q^2}\overline{g\left(\frac{\bx}{\bq}\right)}g(\bx)
e^{-\mathrm{i}\mu \left(\arg\left(\frac{\bx}{\bq}\right)-\arg\left(\bx)\right)\right)}\\
&= e^{-\mathrm{i}\mu \arg(\bq)}\int_{\R^2_{*}}\ud^{2}\bx\,e^{\mathrm{i}\bp\cdot\bx} \,\frac{1}{q^2}\,g\left(\frac{\bx}{\bq}\right)\,g(\bx)\equiv e^{-\mathrm{i}\mu \arg(\bq)}\varpi^{g}(\bq,\bp)
\end{align}
In this case, the related $\Omega^{\psi}$ is given by 
\begin{equation}
\Omega^{\psi}(\bq)= e^{-\mathrm{i}\mu \arg(\bq)}\int_{\R^2_{*}}\ud^{2}\bx \,\frac{1}{q^2}g\left(\frac{\bx}{\bq}\right)(\bx)g(\bx)
= e^{-\mathrm{i}\mu \arg(\bq)} \Omega^{g}(\bq)\, ,
\end{equation}
and leads to $ \partial_{2} \Omega^{\psi}(1) =\ii\mu \Omega^{g}(1)$ and the non-zero flux 
\begin{equation}
\label{psiflux}
\Phi^\vap_0 =  \frac{2\pi\hbar \mu \Omega^{g}(1)}{\mathfrak{q}}\,. 
\end{equation}
Now we combine our results to arrive at 
\begin{equation}
  \frac{(\pmb{\nabla} \Om^{\psi})(1)}{\Omega^{\psi}(1)}= 
    \begin{bmatrix}
        -2 \\ \ii\mu
    \end{bmatrix}\,, \quad    \Delta \Omega^{\psi}(1) = -\mu^2\Vert g\Vert^2+8\pi\left\Vert\frac{1}{Q}g\right\Vert^2+2\pi\left[\Vert g_1^\prime\Vert^2+\Vert g_2^\prime\Vert^2\right]\,.
\end{equation}
We then obtain for the strength of the affine scalar potential:
\begin{equation}
    K^{\varpi_\psi} 
    = 4+2\mu^2+2\pi\frac{\left[\Vert g_1^\prime\Vert^2+\Vert g_2^\prime\Vert^2\right] -4\left\Vert\frac{1}{Q}g\right\Vert^2}{\Vert g\Vert^2}\,.
\end{equation}
By choosing suitable $\mu$ and $g$, we can make this strength positive and as large as  we wish.

\section{Formulae}\label{app:formulae}

For an arbitrary function $F$ with the argument ${\bf x}/{\bf x}^\prime$, 
 the following formulae hold:
\begin{align}
\partial^2_{x^\prime_1} F\left( \frac{\bf x}{\bf x^\prime} \right)
&=
 \left[
\frac{ - 2 x^{\prime}_1 x_1 - 2 x^\prime_2 x_2  }{x^{\prime 4}}
-2 \frac{2x^\prime_1 }{x^{\prime 6}} (- x^{\prime 2}_1 x_1 - 2 x^{\prime}_1 x^\prime_2 x_2 + x_1 x^{\prime 2}_2)
\right] \left. \partial_{u_1} F ({\bf u}) \right|_{{\bf u} = {\bf x}/{\bf x}^\prime} 
\nonumber \\
& + 
\left[
\frac{ - x^{\prime 2}_1 x_1 - 2 x^{\prime}_1 x^\prime_2 x_2 + x_1 x^{\prime 2}_2 }{x^{\prime 4}}
\right] \nonumber \\
&\times \left\{
\left[
\frac{x_1}{x^{\prime 2}} - \frac{2x^\prime_1 (x^\prime_1 x_1 + x^\prime_2 x_2)}{x^{\prime 4}}
\right] \left. \partial^2_{u_1} F ({\bf u}) \right|_{{\bf u} = {\bf x}/{\bf x}^\prime} 
\right. \nonumber \\
& \left. + 
\left[
\frac{x_2}{x^{\prime 2}} - \frac{2x^\prime_1 (x^\prime_1 x_2 - x^\prime_2 x_1)}{x^{\prime 4}}
\right] \left. \partial_{u_1} \partial_{u_2} F ({\bf u}) \right|_{{\bf u} = {\bf x}/{\bf x}^\prime} 
\right\}
\nonumber \\
&
+ 
\left[
\frac{- 2 x^{\prime}_1 x_2 + 2 x^\prime_2 x_1 }{x^{\prime 4}}
- 2 \frac{2x^\prime_1}{x^{\prime 6}}(- x^{\prime 2}_1 x_2 + 2 x^\prime_1 x^\prime_2 x_1 + x^{\prime 2}_2 x_2)
\right]\left. \partial_{u_2} F ({\bf u}) \right|_{{\bf u} = {\bf x}/{\bf x}^\prime} \nonumber \\
&
+ 
\left[
\frac{- x^{\prime 2}_1 x_2 + 2 x^\prime_1 x^\prime_2 x_1 + x^{\prime 2}_2 x_2 }{x^{\prime 4}}
\right] 
\nonumber \\
& \times 
\left\{
\left[
\frac{x_1}{x^{\prime 2}} - \frac{2x^\prime_1 (x^\prime_1 x_1 + x^\prime_2 x_2)}{x^{\prime 4}}
\right] \left. \partial_{u_1} \partial_{u_2} F ({\bf u}) \right|_{{\bf u} = {\bf x}/{\bf x}^\prime} 
\right. \nonumber \\
& \left. + 
\left[
\frac{x_2}{x^{\prime 2}} - \frac{2x^\prime_1 (x^\prime_1 x_2 - x^\prime_2 x_1)}{x^{\prime 4}}
\right] \left. \partial^2_{u_2} F ({\bf u}) \right|_{{\bf u} = {\bf x}/{\bf x}^\prime} 
\right\} \, ,
\end{align}
and
\begin{align}
\partial^2_{x^\prime_2} F\left( \frac{\bf x}{\bf x^\prime} \right)
&= 
 \left[
\frac{ - 2 x^{\prime}_2 x_2 - 2 x^\prime_1 x_1  }{x^{\prime 4}}
-2 \frac{2x^\prime_2 }{x^{\prime 6}} 
(x^{\prime 2}_1 x_2 - 2 x^{\prime}_1 x^\prime_2 x_1 - x_2 x^{\prime 2}_2)
\right] \left. \partial_{u_1} F ({\bf u}) \right|_{{\bf u} = {\bf x}/{\bf x}^\prime} 
\nonumber \\
& + 
\left[
\frac{ x^{\prime 2}_1 x_2 - 2 x^{\prime}_1 x^\prime_2 x_1 - x_2 x^{\prime 2}_2 }{x^{\prime 4}}
\right] 
\nonumber \\
& \times
\left\{
\left[
\frac{x_2}{x^{\prime 2}} - \frac{2x^\prime_2 (x^\prime_1 x_1 + x^\prime_2 x_2)}{x^{\prime 4}}
\right] \left. \partial^2_{u_1} F ({\bf u}) \right|_{{\bf u} = {\bf x}/{\bf x}^\prime} 
\right. \nonumber \\
& \left. + 
\left[
-\frac{x_1}{x^{\prime 2}} - \frac{2x^\prime_2 (x^\prime_1 x_2 - x^\prime_2 x_1)}{x^{\prime 4}}
\right] \left. \partial_{u_1} \partial_{u_2} F ({\bf u}) \right|_{{\bf u} = {\bf x}/{\bf x}^\prime}
\right\}
\nonumber \\ 
& 
+ 
\left[
\frac{- 2 x^{\prime}_1 x_2 + 2 x^\prime_2 x_1 }{x^{\prime 4}}
- 2 \frac{2x^\prime_2}{x^{\prime 6}}
(- x^{\prime 2}_1 x_1 - 2 x^\prime_1 x^\prime_2 x_2 + x^{\prime 2}_2 x_1)
\right]\left. \partial_{u_2} F ({\bf u}) \right|_{{\bf u} = {\bf x}/{\bf x}^\prime} \nonumber \\
& 
+ 
\left[
\frac{- x^{\prime 2}_1 x_1 - 2 x^\prime_1 x^\prime_2 x_2 + x^{\prime 2}_2 x_1 }{x^{\prime 4}}
\right] 
\nonumber \\
& \times 
\left\{
\left[
\frac{x_2}{x^{\prime 2}} - \frac{2x^\prime_2 (x^\prime_1 x_1 + x^\prime_2 x_2)}{x^{\prime 4}}
\right] \left. \partial_{u_1} \partial_{u_2} F ({\bf u}) \right|_{{\bf u} = {\bf x}/{\bf x}^\prime} 
\right. \nonumber \\
& \left. + 
\left[
-\frac{x_1}{x^{\prime 2}} - \frac{2x^\prime_2 (x^\prime_1 x_2 - x^\prime_2 x_1)}{x^{\prime 4}}
\right] \left. \partial^2_{u_2} F ({\bf u}) \right|_{{\bf u} = {\bf x}/{\bf x}^\prime}
\right\} \, .
\end{align}
Using these formulae, we find
\begin{align}
\left\{ \partial_{x^\prime_1} \Omega \left( \frac{\bf Q}{\bf x^\prime}\right)
\right\}_{{\bf x}^\prime = {\bf Q}} 
&= 
- \frac{Q_1}{Q^2}\partial_{1} \Omega (1)
+ 
\frac{Q_2}{Q^{2}} \partial_{2}\Omega (1) \, ,\\
\left\{ 
\partial^2_{x^\prime_1} \Omega \left( \frac{\bf Q}{\bf x^\prime}\right)
\right\}_{{\bf x}^\prime = {\bf Q}} 
&= 
 \left[
-\frac{2}{Q^2} 
+ 4 \frac{Q^2_1 }{Q^4}
\right] \partial_{1} \Omega (1) 
- 4 \frac{Q_1 Q_2}{Q^4}
\partial_2 \Omega (1)
\nonumber \\
& 
+ 
\frac{Q^2_1}{Q^4} \partial^2_{1} \Omega (1) 
 - 2 \frac{Q_1Q_2}{Q^4}
\partial_{1} \partial_{2} \Omega (1)  
+ 
\frac{Q^2_2}{Q^4} \partial^2_{2} \Omega (1)  \, ,\\
\left\{ \partial_{x^\prime_2} \Omega \left( \frac{\bf Q}{\bf x^\prime}\right)
\right\}_{{\bf x}^\prime = {\bf Q}} 
&= 
- \frac{Q_2}{Q^2}  \partial_{1} \Omega (1)  
-\frac{Q_1}{Q^2}  \partial_{2} \Omega (1) \, , \\
\left\{ 
\partial^2_{x^\prime_2} \Omega \left( \frac{\bf Q}{\bf x^\prime}\right)
\right\}_{{\bf x}^\prime = {\bf Q}} 
&= 
 \left[
-\frac{2}{Q^2}
+ 4 \frac{Q^2_2 }{Q^4} 
\right] \partial_{1} \Omega (1)
+ 4 \frac{Q_1Q_2}{Q^4}
 \partial_{2} \Omega (1) 
\nonumber \\
& + 
\frac{Q^2_2}{Q^4} 
\partial^2_{1} \Omega (1)
+ 2 \frac{Q_1 Q_2}{Q^4}  \partial_{1} \partial_{2} \Omega (1)
+\frac{Q^2_1}{Q^4}\partial^2_{2} \Omega (1) \, ,
\end{align}
leading to 
\begin{align}
\left\{ \nabla_{x^\prime} \Omega \left( \frac{\bf Q}{\bf x^\prime}\right)
\right\}_{{\bf x}^\prime = {\bf Q}} \cdot {\bf Q}
&=
-\partial_1 \Omega (1) 
\, , \\
\left\{ \nabla_{x^\prime} \Omega \left( \frac{\bf Q}{\bf x^\prime}\right)
\right\}_{{\bf x}^\prime = {\bf Q}} \cdot {\bf P}
&= 
-\frac{ {\bf Q}}{Q^2} \cdot{\bf P}\partial_1 \Omega (1) + \frac{1}{Q^2}( Q_2P_1 - Q_1 P_2 ) \partial_2 \Omega (1)
\, , \\
\left\{ 
\nabla^2_{x^\prime} \Omega_0 \left( \frac{\bf Q}{\bf x^\prime}\right)
\right\}_{{\bf x}^\prime = {\bf Q}} 
&= 
\frac{1}{Q^2} \nabla^2 \Omega (1) \, .
\end{align}

\end{document}